\documentstyle[12pt,epsfig]{article}
\begin{document}
\noindent
\begin{center}
{\Large {\bf Crossing Phantom Boundary in $f(R)$ Modified Gravity : Jordan Frame vs Einstein Frame}}\\ \vspace{2cm}
 ${\bf Yousef~Bisabr}$\footnote{e-mail:~y-bisabr@srttu.edu.}\\
\vspace{.5cm} {\small{Department of Physics, Shahid Rajaee Teacher
Training University,
Lavizan, Tehran 16788, Iran}}\\
\end{center}
\vspace{1cm}
\begin{abstract}
We study capability of $f(R)$ gravity models to allow crossing the phantom boundary in both Jordan and Einstein conformal
frames.  In Einstein frame, these models
are equivalent to Einstein gravity together with a scalar field minimally
coupled to gravity.  This scalar degree of freedom appears as a quintessence field with a coupling with the matter sector.
We investigate evolution of the equation of sate parameter for some cosmologically viable $f(R)$ gravity models in both conformal frames. This investigation (beyond mere theoretical arguments) acts as an operational tool
to distinguish physical status of the two conformal frames.  It shows that the two conformal frames have not the same physical status.

\end{abstract}
PACS Numbers: 98.80.-k \vspace{3cm}
\section{Introduction}
For explaining accelerated expansion of the universe, there is a class of models in which one modifies the
laws of gravity whereby a late-time acceleration is produced. A
family of these modified gravity models is obtained by replacing
the Ricci scalar $R$ in the usual Einstein-Hilbert Lagrangian
density for some functions $f(R)$ \cite{carro} \cite{sm}.
It is well-known that \cite{maeda} \cite{soko} the gravitational field equations derived from such fourth order gravity
models can be conformally transformed to an Einstein frame representation with an extra scalar degree of freedom.  This
scalar degree of freedom may be regarded as a manifestation of the additional degree of freedom due to the higher order of the
field equations in the Jordan frame.  This feature opens some questions about a possible correspondence between this geometric scalar
field and the usual scalar fields such as quintessence and phantom fields.  Along this line, we intend in the
present contribution to explore phantom behavior of some $f(R)$ gravity models in both Jordan and Einstein
conformal frames.  In particular, we shall show that behaviors of the scalar field
attributed to $f(R)$ models are similar to those of a coupled quintessence, a quintessence which
interacts with the matter sector. \\
The plan of our paper is as follows : In section 2, we consider the cosmology of a scalar field with minimal coupling to gravity.  We review
the observation that a quintessence field with a minimal coupling can not lead to crossing the phantom boundary.
In section 3, we shall consider $f(R)$ gravity in both Jordan and Einstein conformal frames.  We show that the generic feature
of these models is that the scalar partner of the
metric tensor in the Einstein frame is similar to a quintessence rather than a phantom field.  The interaction
of this scalar degree of freedom with the matter sector plays a key role for crossing the phantom barrier \cite{bis}. We then compare phantom behavior of some
viable $f(R)$ models in the two conformal frames.  Our results are summarized in section 4.

~~~~~~~~~~~~~~~~~~~~~~~~~~~~~~~~~~~~~~~~~~~~~~~~~~~~~~~~~~~~~~~~~~~~~~~~~~~~~~~~~~~~~~~~~
\section{Minimally coupled Scalar Field}

The simplest class of models that provides a redshift dependent
equation of state parameter is a scalar field $\varphi$ minimally coupled to
gravity whose dynamics is determined by a properly chosen
potential function $V(\varphi)$.  Such models are described by the
action \footnote{We work in units in which $\hbar=c=1$ and the signature is $(-,+,+,+)$.}
\begin{equation}
S_{\varphi}=\frac{1}{2} \int d^4 x\sqrt{-g}~(\frac{1}{k}R-\alpha ~g^{\mu\nu}\partial_{\mu}\varphi
\partial_{\nu}\varphi-2V(\varphi))+S_{m}(g_{\mu\nu}, \psi)
\label{ca1}\end{equation} where $k=8\pi G$ with $G$ being the gravitational constant, $g$ is the
determinant of $g_{\mu\nu}$ and $R$ is the curvature scalar. Here $S_{m}$ is the action
of dark matter which depends on the metric $g_{\mu\nu}$ and some dark matter
fields $\psi$. The constant $\alpha$ can take $\alpha=+1, -1$ which correspond to quintessence and
phantom fields, respectively. The distinguished feature of the phantom
field is that its kinetic term enters (\ref{ca1}) with opposite
sign in contrast to the quintessence or ordinary matter. The action (\ref{ca1}) gives the Einstein field equations
\begin{equation}
G_{\mu\nu}=k (T_{\mu\nu}^{\varphi}+T_{\mu\nu}^{m})
\label{ca2}\end{equation} with
\begin{equation}
T_{\mu\nu}^{\varphi}=\alpha~\nabla_{\mu}\varphi
\nabla_{\nu}\varphi-\frac{1}{2}\alpha~g_{\mu\nu}
\nabla_{\gamma}\varphi \nabla^{\gamma}\varphi-g_{\mu\nu} V(\varphi)
\label{ca3}\end{equation}
Here $T_{\mu\nu}^m$ is the stress-tensor of the matter system defined by
\begin{equation}
T_{\mu\nu}^m=\frac{-2}{\sqrt{-g}}\frac{\delta S_{m}(g_{\mu\nu},\psi)}{\delta g^{\mu\nu}}
\label{a3}\end{equation}
The two stress-tensors $T_{\mu\nu}^{\varphi}$ and $T_{\mu\nu}^{m}$ are separately conserved
\begin{equation}
\nabla^{\mu}T_{\mu\nu}^{\varphi}=\nabla^{\mu}T_{\mu\nu}^{m}=0
\label{aa3}\end{equation}
We assume a spatially flat homogeneous and isotropic cosmology described by
Friedmann-Robertson-Walker (FRW) spacetime
\begin{equation}
ds^2=-dt^2+a^2(t)(dx^2+dy^2+dz^2)
\label{aa7}\end{equation}
where $a(t)$ is the scale factor. In this cosmology, the gravitational field equations (\ref{ca2}) become
\begin{equation}
3H^2=k(\rho_{\varphi}+\rho_{m})
\label{a08}\end{equation}
\begin{equation}
2\dot{H}=-k[(\omega_{\varphi}+1)\rho_{\varphi}+\rho_{m}]
\label{a09}\end{equation}
where $H\equiv \frac{\dot{a}}{a}$ is the Hubble parameter and
\begin{equation}
\rho_{\varphi}=\frac{1}{2}\alpha
\dot{\varphi}^2+V(\varphi)~,~~~~~p_{\varphi}=\frac{1}{2}\alpha
\dot{\varphi}^2-V(\varphi) \label{ca4}\end{equation}
\begin{equation}
\omega_{\varphi}=\frac{\frac{1}{2}\alpha
\dot{\varphi}^2-V(\varphi)}{\frac{1}{2}\alpha
\dot{\varphi}^2+V(\varphi)} \label{ca5}\end{equation}
The conservation equations (\ref{aa3}) take the form
\begin{equation}
\dot{\rho}_m+3H\rho_m=0
\label{ab1}\end{equation}
\begin{equation}
\dot{\rho}_{\varphi}+3H(\omega_{\varphi}+1)\rho_{\varphi}=0
\label{ab2}\end{equation}
 In the case
of a quintessence field ($\alpha=+1$) with $V(\varphi)>0$
the equation of state parameter remains in the
range $-1<\omega_{\varphi}<1$.  This is also true for a phantom field ($\alpha=-1$) with a negative potential $V(\varphi)<0$.  In the limit of small kinetic term
(slow-roll potentials \cite{slow}), it approaches
$\omega_{\varphi}=-1$ but does not cross this line.  For $\alpha=+1$, we have
\begin{equation}
\rho_{\varphi}+p_{\varphi}=(\omega_{\varphi}+1)\rho_{\varphi}=\alpha \dot{\varphi}^2>0
\label{wec}\end{equation}
The phantom
barrier can be crossed by a phantom field ($\alpha<0$) with
$V(\varphi)>0$ when we have $2|V(\varphi)|>\dot{\varphi}^2$.  This
situation corresponds to
\begin{equation}
\rho_{\varphi}>0~~~~~,~~~~~p_{\varphi}<0~~~~~,~~~~~V(\varphi)>0
\label{a51}\end{equation}
In this case, $(\omega_{\varphi}+1)\rho_{\varphi}<0$ and for a sufficiently negative pressure $p_{\varphi}$ the equation
(\ref{a09}) gives $\dot{H}>0$.  Let us look at this situation in a different way.  We combine (\ref{a09}) with (\ref{ab1}) and (\ref{ab2}) which
leads to
\begin{equation}
2\dot{H}=\frac{k}{3H}(\dot{\rho}_{\varphi}+\dot{\rho}_{m})
\label{a19}\end{equation}
In an expanding universe $\dot{\rho}_m<0$.  In the case of a quintessence $(\omega_{\varphi}+1)\rho_{\varphi}>0$ and then the equation
(\ref{ab2}) implies that $\dot{\rho}_{\varphi}<0$ which leads to $\dot{H}<0$.  On
the other hand, for a phantom field we have $\alpha<0$
and $(\omega_{\varphi}+1)\rho_{\varphi}<0$.  In this case, $\dot{\rho}_{\varphi}>0$ and for an appropriate potential
function $V(\varphi)$ the first term on the right hand side of the equation (\ref{a19}) dominates and the equation
then results in $\dot{H}>0$, crossing the phantom barrier.  The merit of (\ref{a19}) is that it implicitly shows that there is
a possibility for an interacting
quintessence field to cross the phantom boundary. We will return to this issue in the next section.\\
Here it is assumed that the scalar field has a canonical kinetic
term $\pm \frac{1}{2}\dot{\varphi}^2$. It is shown \cite{vik} that
any minimally coupled scalar field with a generalized kinetic term
(k-essence Lagrangian \cite{k}) can not lead to crossing the PDL
through a stable trajectory.  However, there are models that
employ Lagrangians containing multiple fields \cite{multi} or
scalar fields with non-minimal coupling \cite{non} which in
principle can achieve crossing the barrier.
~~~~~~~~~~~~~~~~~~~~~~~~~~~~~~~~~~~~~~~~~~~~~~~~~~~~~~~~~~~~~~~~~~~~~~~~~~~~~~~~~~~~~~~~~~

\section{f(R) Gravity}
The $f(R)$ gravity models are based on a Lagrangian density which depends in a nonlinear way on the curvature scalar.
In these models the dynamical variable of the vacuum sector
is the metric tensor and the corresponding field equations are fourth order.  This
dynamical variable can be replaced
by a new pair which consists of a conformally rescaled metric and a scalar partner.  Moreover, in terms of the new set
of variables the field equations are those of General Relativity.  The original set of variables
is commonly called Jordan conformal frame and the transformed set whose dynamics is described by Einstein field equations
is called Einstein conformal frame.\\
 In general, the mathematical equivalence of these two conformal frames does not imply a physical equivalence.  The physical status
of these conformal frames is an open question which has not been completely solved yet.  In fact, there
are different categories according to various authors' attitude towards the issue of which frame is the physical one \cite{soko} \cite{fa}. Based on the interaction of the scalar degree of freedom with the matter sector, some authors argue that
the two conformal frames are physically equivalent provided that one accepts the idea that the units of mass, length and time
are varying in the Einstein frame \cite{dick}.  According to this idea, physics must be conformally invariant and symmetry
group of gravitational theories should be enlarged to include not only the group of diffeomorphisms but also conformal transformations. There
are also arguments against this idea and in favor of one of the conformal frames.  In particular, some authors go beyond
mere theoretical arguments and take an appropriate phenomenon to determine if the two frames are physically equivalent \cite{20}.\\
One important issue in the context of cosmological viability of an $f(R)$ gravity model is capability of the model
to exhibit phantom behavior due to curvature corrections.  We shall explore this issue in both Jordan and Einstein conformal frames
and take it as a relevant issue to distinguish the physical status of the two conformal frames.
\subsection{Jordan Frame Representation}
The action for an $f(R)$
gravity theory in the Jordan frame is given by
\begin{equation}
S_{JF}=\frac{1}{2 k} \int d^{4}x \sqrt{-g} f(R) +S_{m}(g_{\mu\nu}, \psi)
\label{a1}\end{equation}
Stability issues should be
considered to make sure that an $f(R)$ model is viable \cite{st}.
In particular, stability in matter sector (the Dolgov-Kawasaki
instability \cite{dk}) imposes some conditions on the functional
form of $f(R)$ models.  These
conditions require that the first and the second derivatives of
$f(R)$ functions with respect to the Ricci scalar $R$ should be
positive definite.  The positivity of the first derivative ensures
that the scalar degree of freedom is not tachyonic and positivity
of the second derivative tells us that graviton is not a ghost.\\
The field equations can be derived by varying the action with respect to the metric tensor
\begin{equation}
f'(R) R_{\mu\nu}-\frac{1}{2}g_{\mu\nu}f(R)-\nabla_{\mu}\nabla_{\nu}f'(R)+g_{\mu\nu} \Box f'(R)=k T_{\mu\nu}^m
\label{a2}\end{equation}
where prime denotes differentiation with respect to $R$.  The trace of (\ref{a2}) is
\begin{equation}
f'(R) R-2f(R)+3\Box f'(R)=k T^{m}
\label{a4}\end{equation}
where $T^m=g^{\mu\nu}T^m_{\mu\nu}$.  It should be noted that when $f(R)=R$,  the equation (\ref{a4}) reduces to $R=-kT^m$ which is the corresponding trace equation in General Relativity.  In $f(R)$ gravity, $\Box f'(R)$ does not vanish and contrary to General Relativity the Ricci scalar relates
differentially to the trace of the matter system.  This is an indication of the fact that one finds a larger variety of solutions in $f(R)$
modified gravity models.\\
The field equations can also be written in the following form
\begin{equation}
G_{\mu\nu}=k_{eff}(T_{\mu\nu}^m +T_{\mu\nu}^{DE})
\label{a5}\end{equation}
where $k_{eff}=k/f'(R)$ and
\begin{equation}
T_{\mu\nu}^{DE}=\frac{1}{k}[\frac{1}{2}(f(R)-Rf'(R))g_{\mu\nu}+\nabla_{\mu}\nabla_{\nu}f'(R)-g_{\mu\nu}\Box f'(R)]
\label{a6}\end{equation}
We apply the field equations to a
spatially flat FRW spacetime described by (\ref{aa7}).
We assume that the matter system is described by a pressureless perfect fluid with energy
density $\rho_{m}$.  The field equations become
\begin{equation}
3H^2=k_{eff}(\rho_m+\rho_{DE})
\label{a8}\end{equation}
\begin{equation}
2\dot{H}+3H^2=-k_{eff}~p_{DE}
\label{a9}\end{equation}
where
\begin{equation}
\rho_{DE}=\frac{1}{k}[\frac{1}{2}(Rf'(R)-f(R))-3H\dot{R}f''(R)]
\label{a10}\end{equation}
\begin{equation}
p_{DE}=\frac{1}{k}[\frac{1}{2}(f(R)-Rf'(R))+\ddot{R}f''(R)+\dot{R}^2 f'''(R)+2H \dot{R}f''(R)]
\label{a11}\end{equation}
Then the effective equation of state parameter is
\begin{equation}
\omega_{eff}=\frac{p_{DE}}{\rho_{m}+\rho_{DE}}=\frac{1}{3H^2f'(R)}[\frac{1}{2}(f(R)-R f'(R))+\ddot{R}f''(R)+\dot{R}^2 f'''(R)+2H \dot{R}f''(R)]
\label{a12}\end{equation}
where we have used (\ref{a8}).  This expression allows us to find those $f(R)$ functions that fulfill $\omega_{eff}<-1$. In general, to find
such $f(R)$ gravity models one may start with a particular $f(R)$
function in the action (\ref{a1}) and solve the corresponding
field equations for finding the form of $H(z)$.  One can then use
this function in (\ref{a12}) to obtain $\omega_{eff}(z)$.
However, this approach is not efficient in view of
complexity of the field equations.  An alternative
approach is to start from the best fit parametrization $H(z)$
obtained directly from data and use this $H(z)$ for a particular
$f(R)$ function in (\ref{a12}) to find $\omega_{eff}(z)$.  In this approach, one needs a consistency check to ensure that the given parametrization
 is compatible with the $f(R)$ function.  We
will follow the latter approach to find $f(R)$ models that
provide crossing the phantom barrier.  We begin with the Hubble parameter $H$ which its derivative with respect to cosmic time $t$ is
\begin{equation}
\dot{H}=\frac{\ddot{a}}{a}-(\frac{\dot{a}}{a})^2
\label{b11}\end{equation}
Combining this with the
definition of the deceleration parameter
\begin{equation}
q(t)=-\frac{\ddot{a}}{aH^2} \label{b12}\end{equation} gives
\begin{equation}
\dot{H}=-(q+1)H^2 \label{bb13}\end{equation} One may use
$z=\frac{a(t_{0})}{a(t)}-1$ with $z$ being the redshift, and the
relation (\ref{b12}) to write (\ref{bb13}) in its integration form
\begin{equation}
H(z)=H_{0}~\exp[\int_{0}^{z} (1+q(u))d\ln(1+u)]
\label{bc14}\end{equation} where the subscript ``0" indicates the
present value of a quantity.  Now if a function $q(z)$ is given,
then we can find evolution of the Hubble parameter. Here we use a
two-parametric reconstruction function characterizing $q(z)$
\cite{wang}\cite{q},
\begin{equation}
q(z)=\frac{1}{2}+\frac{q_{1}z+q_{2}}{(1+z)^2}
\label{bc15}\end{equation}
One of the advantages of this parametrization is that $q(z)\rightarrow \frac{1}{2}$ when $z \gg 1$ which is consistent with observations.
Moreover, The behavior of q(z) in this parametrization is quite
general.
If $q_1 > 0$
and $q_2 > 0$, then there is no acceleration at all. It is also possible that the Universe has been
accelerating since some time in the past. If $q_1 < 0$ and $q_2 > -1/2$, then it is possible that
the Universe is decelerating and has past acceleration and deceleration.  In other words, it has the same behavior
as the simple three-epoch model used in \cite{wang} \cite{shap}.  The values of $q_1$ and
$q_2$ and the behavior of $q(z)$ can be obtained by fitting the model to the observational data.
Fitting this model to the Gold
data set gives $q_{1}=1.47^{+1.89}_{-1.82}$ and $q_{2}=-1.46\pm
0.43$\footnote{Some other recent samples may be found in \cite{sam}.} \cite{q}. Using this in (\ref{bc14}) yields
\begin{equation}
H(z)=H_{0}(1+z)^{3/2}~\exp[\frac{q_{2}}{2}+\frac{q_{1}z^2-q_{2}}{2(z+1)^2}]
\label{bc16}\end{equation} For the metric (\ref{aa7}), we have $R=6(\dot{H}+2H^2)$ and therefore $\dot{R}=6(\ddot{H}+4\dot{H}H)$.
In terms of the deceleration parameter, one can write
\begin{equation}
R=6(1-q)H^2 \label{b17}\label{bc17}\end{equation}and
\begin{equation}
\dot{R}=6H^3 \{2(q^2-1)-\frac{\dot{q}}{H}\}
\label{b18}\end{equation}
\begin{equation}
\ddot{R}=6H^3\{6(q^2-1)\frac{\dot{H}}{H}+4q\dot{q}-2\frac{\dot{q}\dot{H}}{H^2}-\frac{\ddot{q}}{H}\}
\label{bb18}\end{equation}
The latter two equations are equivalent to
\begin{equation}
\dot{R}=6H^3 \{2(q^2-1)+(1+z)\frac{dq}{dz}\}
\label{b19}\end{equation}
\begin{equation}
\ddot{R}=-6(z+1)H^3\{3\frac{dH}{dz}[2(q^2-1)+(z+1)\frac{dq}{dz}]+H[(4q+1)\frac{dq}{dz}+(z+1)\frac{d^2q}{dz^2}]\}
\label{bb19}\end{equation}
From the equations (\ref{bc15}) and (\ref{bc16}) we can also write
\begin{equation}
\frac{dq}{dz}=\frac{(q_{1}-2q_{2})-q_{1}z}{(z+1)^3}
\label{b20}\end{equation}
\begin{equation}
\frac{d^2q}{dz^2}=\frac{2q_{1}(z-2)+6q_{2}}{(z+1)^4}
\label{b21}\end{equation}
\begin{equation}
\frac{dH}{dz}=H_{0}(1+z)^{1/2}[\frac{3}{2}+\frac{q_{1}z+q_2}{(z+1)^2}]~\exp[\frac{q_{2}}{2}+\frac{q_{1}z^2-q_{2}}{2(z+1)^2}]
\label{b22}\end{equation}
It is now possible to find $R$, $\dot{R}$ and $\ddot{R}$ in terms of the
redshift, and then for a given $f(R)$ function, the relation
(\ref{a12}) determines the evolution of the equation of state
parameter $\omega_{eff}(z)$.\\
As an illustration we apply this
procedure to some $f(R)$ functions. Let us first consider the
model \cite{cap} \cite{A}
\begin{equation}
f(R)=R+\lambda R_0(\frac{R}{R_0})^n \label{a13}\end{equation}
Here $R_{0}$ is taken to be of the order of $H_{0}^2$ and $\lambda$, $n$ are constant parameters. In terms of the values
attributed to these parameters, the model
(\ref{a13}) is divided in three cases \cite{A}. Firstly, when
$n>1$ there is a stable matter-dominated era which does not follow
by an asymptotically accelerated regime. In this case, $n = 2$
corresponds to Starobinsky's inflation and the accelerated phase
exists in the asymptotic past rather than in the future. Secondly,
when $0<n<1$ there is a stable matter-dominated era followed by an
accelerated phase only for $\lambda<0$. Finally, in the case that
$n<0$ there is no accelerated and matter-dominated phases for
$\lambda>0$ and $\lambda<0$, respectively.  Thus the model
(\ref{a13}) is cosmologically viable in the regions of the
parameters space which is given by $\lambda<0$ and $0<n<1$.\\
For the model (\ref{a13}), one can write
$$
f'(R)=1+n\lambda(\frac{R}{H_0^2})^{n-1}
$$
\begin{equation}
f''(R)=n(n-1)\lambda H_0^{-2}(\frac{R}{H_0^2})^{n-2}
\label{a14}\end{equation}
$$
f'''(R)=n(n-1)(n-2)\lambda H_0^{-4}(\frac{R}{H_0^2})^{n-3}
$$
Putting these results together with (\ref{bc16}), (\ref{bc17}), (\ref{b19}) and (\ref{bb19}) into (\ref{a12}) gives $\omega_{eff}(z)$.   Due
to complexity of the resulting $\omega_{eff}(z)$  function, we do
not explicitly write it here and only plot it in fig.1a for some
parameters. The figure
indicates that $\omega_{eff}(z)$ crosses the phantom boundary for some values of the parameters.   \\
Now we consider the model presented by Starobinsky  \cite{star}
\begin{equation}
f(R)=R-\gamma R_{0} \{1-[1+(\frac{R}{R_{0}})^2]^{-m}\}
\label{d12}\end{equation}
where $\gamma$, $m$ are
positive constants and $R_{0}$ is again of the order of the
presently observed effective cosmological constant.  For this model, one obtains
$$
f'(R)=1-2m\gamma \frac{R}{H_0^2}(1+\frac{R^2}{H_0^4})^{-m-1}
$$
\begin{equation}
f''(R)=2m\gamma H_0^{-2}(1+\frac{R^2}{H_0^4})^{-m-1} ~[2(m+1)\frac{R^2}{H_0^4}(1+\frac{R^2}{H_0^4})^{-1}-1]
\label{a15}\end{equation}
$$
f'''(R)=4m(m+1)\gamma \frac{R}{H_0^6}(1+\frac{R^2}{H_0^4})^{-m-3}~[3-(2m+1)\frac{R^{2}}{H_0^4}]
$$
The corresponding $\omega_{eff}(z)$ function is plotted in fig.2a.  The figure indicates that the Starobinsky's model
also realizes crossing the phantom boundary.
~~~~~~~~~~~~~~~~~~~~~~~~~~~~~~~~~~~~~~~~~~~~~~~~~~~~~~~~~~~~~~~~~~~~~~~~~
\subsection{Einstein Frame Representation}
Any $f(R)$ gravity model described by the action (\ref{a1}) can be recast by a new
set of variables
\begin{equation}
\bar{g}_{\mu\nu} =\Omega~ g_{\mu\nu} \label{b2}\end{equation}
\begin{equation} \phi = \frac{1}{2\beta \sqrt{k}} \ln \Omega
\label{b3}\end{equation}
 where
$\Omega\equiv f^{'}(R)$ and $\beta=\sqrt{\frac{1}{6}}$.
This is indeed a conformal transformation which transforms the
action (\ref{a1}) to the Einstein frame
\cite{maeda} \cite{soko}
\begin{equation}
S_{EF}=\frac{1}{2} \int d^{4}x \sqrt{-\bar{g}}~\{ \frac{1}{k} \bar{R}-\bar{g}^{\mu\nu}
\partial_{\mu} \phi~ \partial_{\nu} \phi -2V(\phi)\} + S_{m}(\bar{g}_{\mu\nu}
e^{2\beta \sqrt{k}\phi}, \psi) \label{b4}\end{equation} In the Einstein
frame, $\phi$ is a minimally coupled scalar field with a
self-interacting potential which is given by
\begin{equation}
V(\phi(R))=\frac{Rf'(R)-f(R)}{2 k f'^2(R)} \label{b5}\end{equation}
Note that the conformal transformation induces the coupling of the
scalar field $\phi$ with the matter sector. The strength of this
coupling $\beta$, is fixed to be $\sqrt{\frac{1}{6}}$ and is the
same for
all types of matter fields. \\
Variation of the action (\ref{b4}) with respect to
$\bar{g}_{\mu\nu}$ and $\phi$ gives,
\begin{equation}
\bar{G}_{\mu\nu}=k (T^{\phi}_{\mu\nu}+\bar{T}^{m}_{\mu\nu})
\label{b6}\end{equation}
\begin{equation}
\bar{\Box} \phi-\frac{d V}{d \phi}=-\beta \sqrt{k}~ \bar{T}^m
\label{bbb6}\end{equation}
 where
\begin{equation}
\bar{T}^{m}_{\mu\nu}=\frac{-2}{\sqrt{-g}}\frac{\delta
S_{m}(\bar{g}_{\mu\nu}
e^{2\beta \sqrt{k}\phi}, \psi)}{\delta \bar{g}^{\mu\nu}}\label{b7}\end{equation}
\begin{equation}
T^{\phi}_{\mu\nu}=\nabla_{\mu} \phi~\nabla_{\nu} \phi
-\frac{1}{2}\bar{g}_{\mu\nu} \nabla_{\gamma}
\phi~\nabla^{\gamma} \phi-V(\phi) \bar{g}_{\mu\nu}
\label{b8}\end{equation} Here $\bar{T}^{m}_{\mu\nu}$ and
$T^{\phi}_{\mu\nu}$ are stress-tensors of the matter system and
the minimally coupled scalar field $\phi$, respectively.
The trace of (\ref{b6}) is
\begin{equation}
\nabla^{\gamma}\phi
\nabla_{\gamma}\phi+4V(\phi)-\bar{R}/k=\bar{T}^m\label{b8-1}\end{equation}
which differentially relates the trace of the matter stress-tensor $\bar{T}^{m}=\bar{g}^{\mu\nu}\bar{T}^m_{\mu\nu}$
to $\bar{R}$.  It is important to note that the two
stress-tensors $\bar{T}^m_{\mu\nu}$ and $T^{\phi}_{\mu\nu}$
are not separately conserved.
 Instead they satisfy the following equations
\begin{equation}
\bar{\nabla}^{\mu}\bar{T}^{m}_{\mu\nu}=-\bar{\nabla}^{\mu}T^{\phi}_{\mu\nu}= \beta \sqrt{k}~\nabla_{\nu}\phi~\bar{T}^{m}\label{b13}\end{equation}
We apply the field equations to a
spatially flat FRW metric which in the Einstein frame is given by
\begin{equation}
d\bar{s}^2=-d\bar{t}^2+\bar{a}^2(t)(dx^2+dy^2+dz^2)
\label{aaa7}\end{equation}
where $\bar{a}=\Omega^{\frac{1}{2}}a$ and $d\bar{t}=\Omega^{\frac{1}{2}}dt$.  To do this, we take
$\bar{T}^m_{\mu\nu}$ and $T^{\phi}_{\mu\nu}$ as the stress-tensors of a pressureless perfect fluid with energy density
$\bar{\rho}_{m}$, and a perfect fluid with energy density
$\rho_{\phi}=\frac{1}{2}\dot{\phi}^2+V(\phi)$ and pressure
$p_{\phi}=\frac{1}{2}\dot{\phi}^2-V(\phi)$, respectively. In this
case, for the metric (\ref{aaa7}) the equations (\ref{b6}) and (\ref{bbb6}) take the form \footnote{Hereafter we will use unbarred characters in the Einstein frame.}
\begin{equation}
3H^2=k(\rho_{\phi}+\rho_{m})
\label{b14}\end{equation}
\begin{equation}
2\dot{H}=-k[(\omega_{\phi}+1)\rho_{\phi}+\rho_m]
\label{b14-1}\end{equation}
\begin{equation}
\ddot{\phi}+3H\dot{\phi}+\frac{dV(\phi)}{d\phi}=-\beta \sqrt{k}~\rho_{m}
\label{b15}\end{equation}
The trace equation (\ref{b8-1}) and the conservation equations
(\ref{b13}) give, respectively,
\begin{equation}
\dot{\phi}^2+R/k-4V(\phi)=\rho_{m}
\label{b16}\end{equation}
\begin{equation}
\dot{\rho}_{m}+3H\rho_{m}=Q \label{b17}\end{equation}
\begin{equation}
\dot{\rho}_{\phi}+3H(\omega_{\phi}+1)\rho_{\phi}=-Q
\label{b18}\end{equation} where
\begin{equation}
Q=\beta \sqrt{k} \dot{\phi}\rho_{m}
\label{b-18}\end{equation} is the interaction term.  This term
vanishes only for $\phi=const.$, which due to (\ref{b3}) corresponds to the case that
$f(R)$ linearly depends on $R$. The direction of energy
transfer depends on the sign of $Q$ or $\dot{\phi}$.  For
$\dot{\phi}>0$, the energy transfer is from dark energy to dark
matter and for $\dot{\phi}<0$ the reverse is true.\\
The formulation of an $f(R)$ theory in the Einstein frame indicates that such a modified gravity model has one extra scalar degree of freedom compared with General Relativity.  This suggests that it is this scalar
degree of freedom which drives late-time acceleration or crossing the PDL in a cosmologically viable $f(R)$ gravity.  This
opens questions about the role of this scalar degree of freedom as quintessence or phantom.  Comparing the sign of the
kinetic terms of the scalar field in the actions (\ref{b4}) and (\ref{ca1}) immediately reveals that this scalar degree of freedom actually appears as a quintessence rather than a phantom
field. It is well-known that a minimally coupled quintessence field may lead to an accelerated expansion but it can not lead to crossing the PDL. This may lead one to conclude that crossing the PDL can not take place in models such as (\ref{b4}).  However, there is a basic difference between the two actions (\ref{ca1}) and (\ref{b4}).  In the former there is no interaction between the
two fluids while in the latter the scalar field $\phi$ interacts with the matter
sector.  In the following we shall show that this interaction is actually responsible for a possible crossing of the PDL in a viable $f(R)$ model.\\
To do this, we first combine (\ref{b17}) and (\ref{b18}) with (\ref{b14-1}) to obtain
\begin{equation}
2\dot{H}=\frac{k}{3H}(\dot{\rho}_{\phi}+\dot{\rho}_{m})
\label{d1}\end{equation}
Apart from some similarities between the latter and the equation (\ref{a19}), there is an important difference between the two equations concerning the interaction between the scalar field $\phi$ and the matter part.  In general, we can consider two different cases \footnote{The case of $Q=0$ corresponds to $\Lambda$CDM.  In this case, the equation (\ref{b17}) states that $\rho_{m}$ always decreases
with expansion of the universe.} :\\
a) Firstly, $Q>0$ which corresponds to energy transfer from dark energy (or the scalar degree of freedom $\phi$) to (dark) matter.  Due to the fact that $\phi$ appears
in our case as a quintessence, we have always $\rho_{\phi}(\omega_{\phi}+1)>0$ for both signs of $Q$.  Thus in the case that $Q>0$, $\dot{\rho}_{\phi}$ is always a decreasing function in an expanding universe ($H>0$).  From
(\ref{b17}) one infers that $\dot{\rho}_m$ can take both positive and negative signs depending on the relative magnitudes of $Q$ and
$3H\rho_m$.  The equation (\ref{d1}) implies that when $\dot{\rho}_m>0$ one can consider the possibility that $\dot{H}>0$ and crossing
the phantom barrier.\\
b) Secondly, $Q<0$ corresponds to a reversed sign of energy transfer.  In this case $\dot{\rho}_m$ is definitely negative in an
expanding universe.  Instead $\dot{\rho}_{\phi}$ can be negative and positive depending on the relative magnitudes of $\rho_{\phi}(\omega_{\phi}+1)$
and $Q$.  The case that $\dot{\rho}_{\phi}>0$ may lead the accelerating expansion to cross the PDL.\\  This argument emphasizes the role of interaction
in the sign of $\dot{H}$ even though it does not appear explicitly in the equation (\ref{d1}).  The effective equation of
state parameter is defined by
\begin{equation}
\omega_{eff}=\frac{p_{eff}}{\rho_{eff}}=\frac{\omega_{\phi}\rho_{\phi}}{\rho_{\phi}+\rho_m}
\label{d2}\end{equation}
We may write it in a more appropriate form.  To do this, we can use (\ref{b14}), (\ref{b17}) and (\ref{b18}) that leads to
\begin{equation}
\omega_{eff}=-1-\frac{k}{9H^3}(\dot{\rho}_{\phi}+\dot{\rho}_{m})=-1+\frac{k}{3H^2}(\frac{3}{k}H^2+\omega_{\phi}\rho_{\phi})
\label{d4}\end{equation}
Using $p_{\phi}=\omega_{\phi}\rho_{\phi}$, the latter takes the form
\begin{equation}
\omega_{eff}=-1+\frac{k}{3H^2}(\frac{3}{k}H^2+\frac{1}{2}\dot{\phi}^2-V(\phi))
\label{d5}\end{equation}
Substituting (\ref{b3}) and (\ref{b5}) into this equation, we obtain
\begin{equation}
\omega_{eff}=-1+\frac{1}{3H^2}[3H^2-\frac{R}{2f'(R)}+\frac{f(R)}{2f'^2(R)}+\frac{3}{4}(\frac{f''(R)}{f'(R)})^2\dot{R}^2]
\label{d6}\end{equation}
It is now possible to find evolution of the equation of state parameter $\omega_{eff}(z)$ with the same procedure used
in the section 3.1.  In fig.1b and fig.2b, the resulting $\omega_{eff}(z)$ is plotted for the models (\ref{a13}) and (\ref{d12}).
Both figures indicate that there are regions in the parameters spaces for which crossing the phantom barrier is allowed.
~~~~~~~~~~~~~~~~~~~~~~~~~~~~~~~~~~~~~~~~~~~~~~~~~~~~~~~~~~~~~~~~~~~~~~~~~~~~~~~~~~~~~~~~~~
\section{Conclusion}

We have studied phantom behavior for some $f(R)$ gravity models in
both Jordan and Einstein conformal frames.  In Jordan frame, we have used the reconstruction function $q(z)$
fitting to the Gold data set to find evolution of equation of
state parameter for some cosmologically viable
$f(R)$ models.  Both models (\ref{a13}) and (\ref{d12}) indicate phantom behavior in some regions of the parameters space. \\ In Einstein frame, the scalar partner of the metric tensor is separated.
Comparing this scalar field with the usual notion of a phantom field, we have made
an observation that the former appears as a \emph{quintessence} with a minimal coupling to gravity.  However, this quintessence field
interacts with matter sector which allows energy transfer between the two components. This interaction
plays a key role in the cosmological behavior of a particular $f(R)$ model in Einstein conformal frame.  Then we have used the same reconstruction function for $q(z)$ to study the evolution of the
effective equation of state parameter.  We have found that both models (\ref{a13}) and (\ref{d12}) can display crossing the phantom boundary in the
same regions of the parameters spaces that is used in Jordan conformal frame.\\  To compare the behavior of $\omega_{eff}(z)$ for each model in the two conformal frames, this function is plotted
for a common set of the parameters.  From fig.1 and fig.2, one can see that
overall behaviors of the equation of state parameters are very similar.  In particular, both models display crossing
the phantom boundary nearly at the same epoch in both conformal frames.  Despite this similarity,
details of the equation of state parameters in the two frames are not exactly the same.
This result stands against the idea (which is partly presented in \cite{fla}) that considers both conformal frames with the same physical status.

\vspace{2cm}
{\bf Figures :}\\\\

\begin{figure}[ht]
\begin{center}
\includegraphics[width=0.45\linewidth]{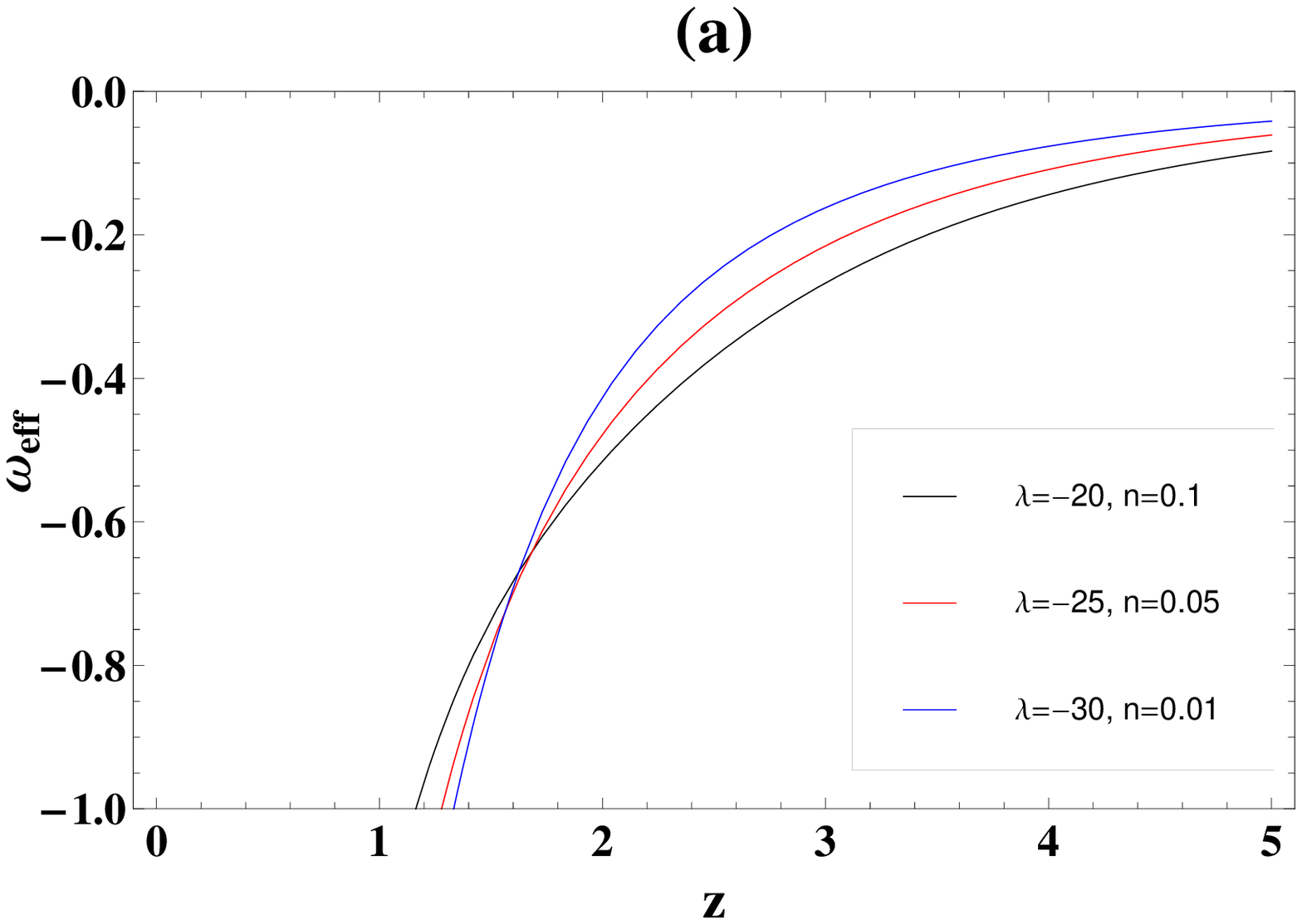}
\includegraphics[width=0.45\linewidth]{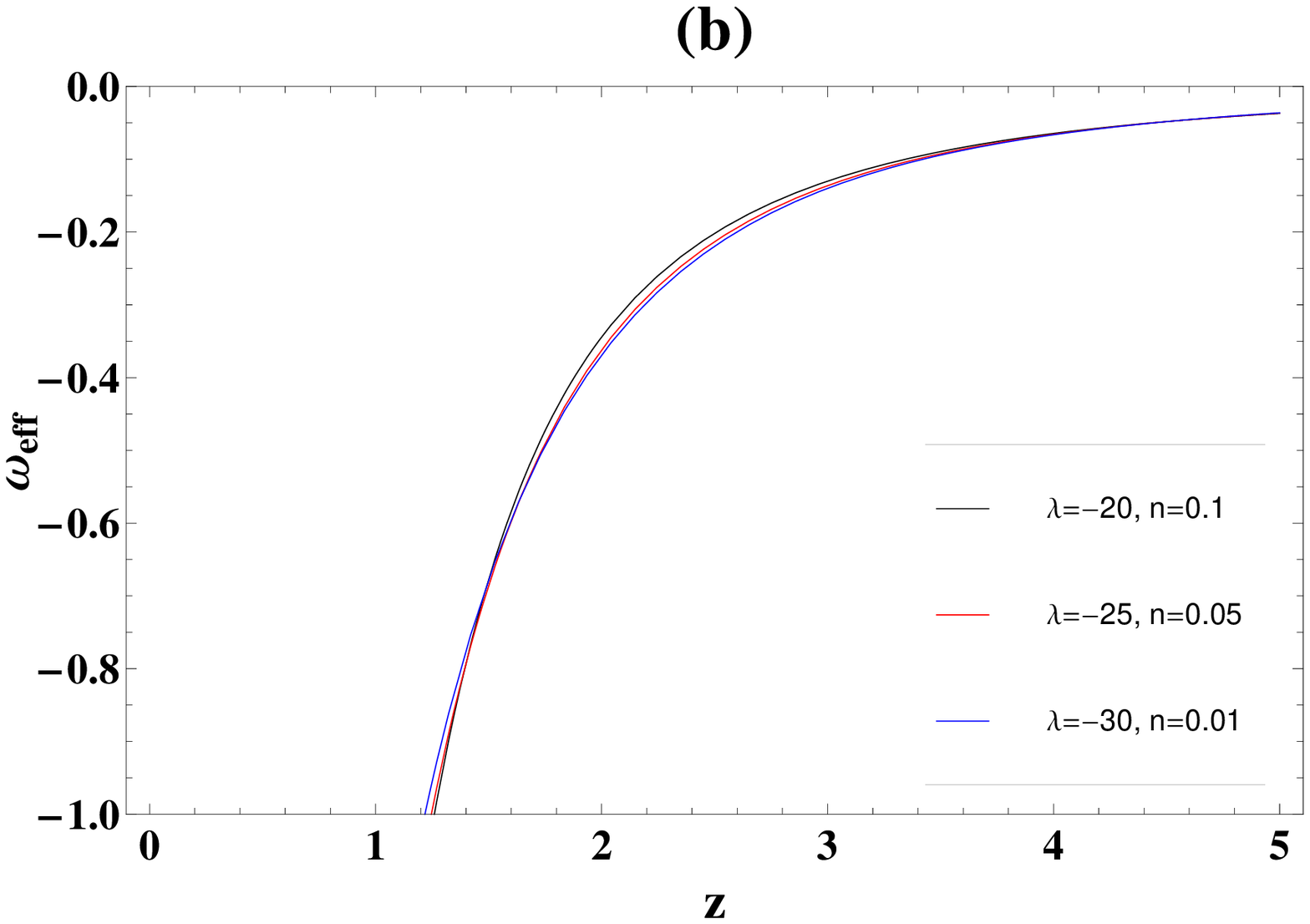}
\caption{The plot of $\omega_{eff}(z)$ for the model (\ref{a13}) in a) Jordan and b) Einstein frames.  The function $\omega_{eff}(z)$ is plotted for a common set of the parameters $\lambda$ and $n$.  The figures show that crossing the PDL
can be realized in both conformal frames.}
\end{center}
\end{figure}

\begin{figure}[ht]
\begin{center}
\includegraphics[width=0.45\linewidth]{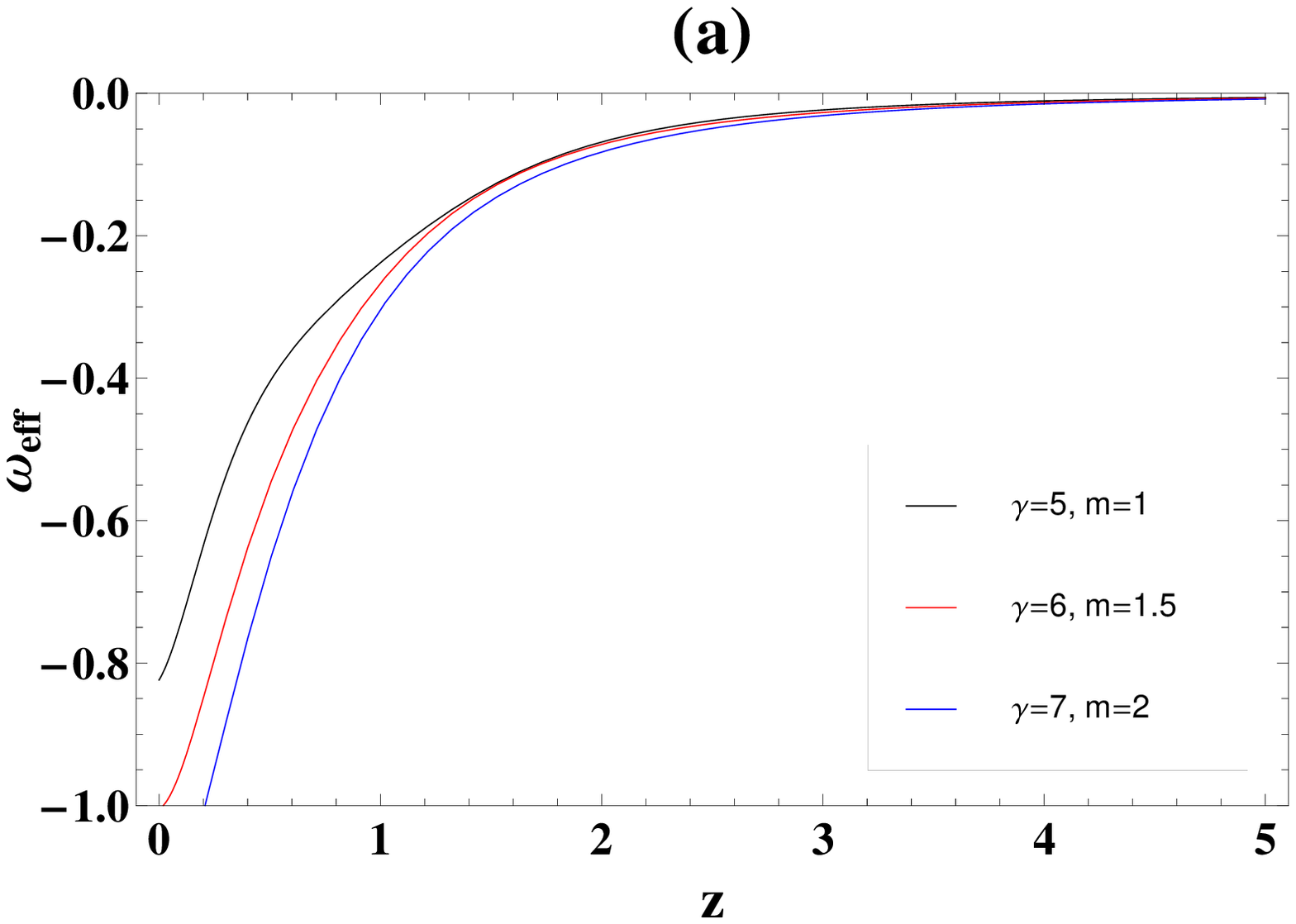}
\includegraphics[width=0.45\linewidth]{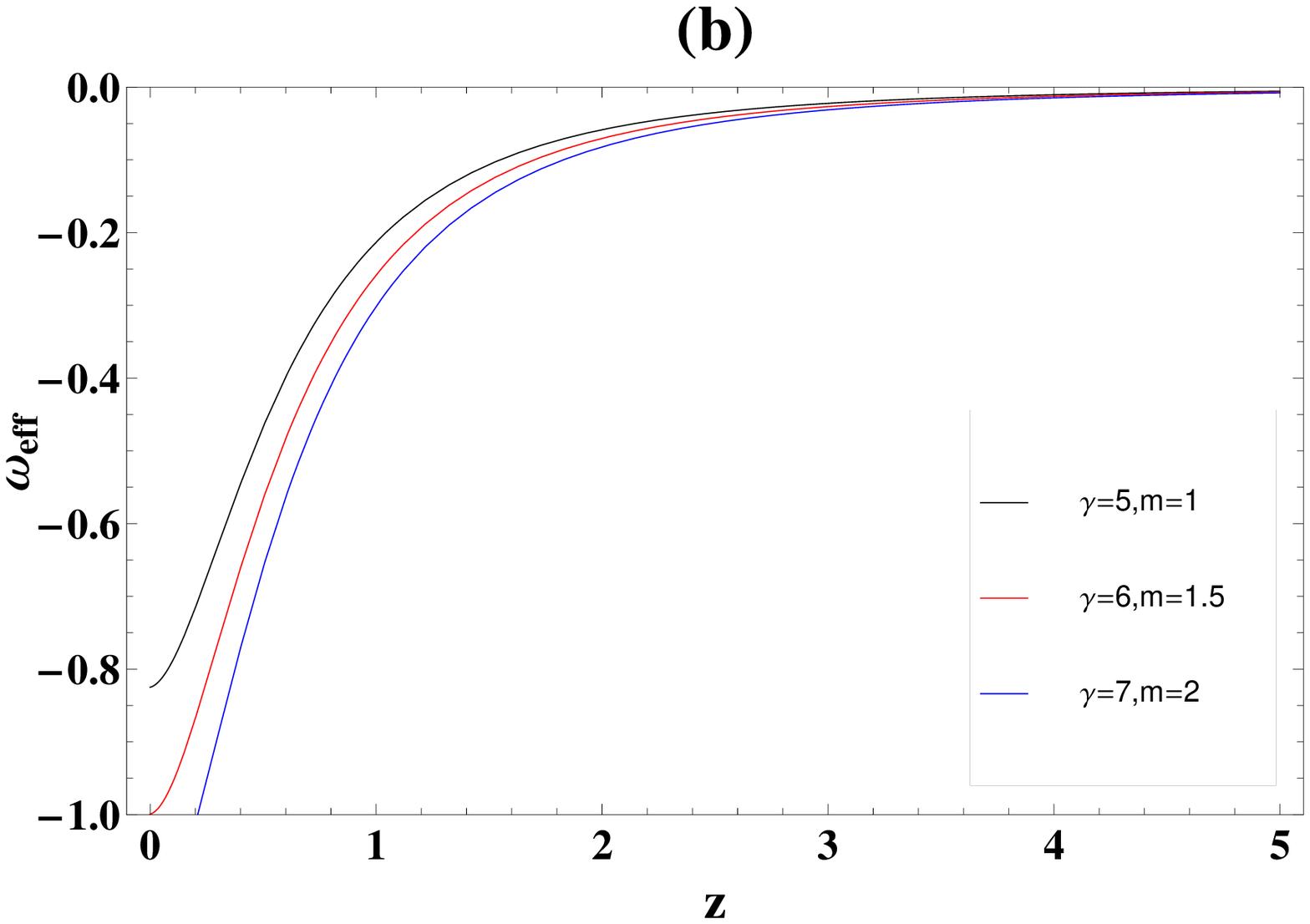}
\caption{The plot of $\omega_{eff}(z)$ for the model (\ref{d12}) in a) Jordan and b) Einstein frames.  The function $\omega_{eff}(z)$ is plotted for a common set of the parameters $\gamma$ and $m$.  The figures show that the PDL
can be crossed in both conformal frames.}
\end{center}
\end{figure}


\begin{thebibliography}{99}

\bibitem{carro}S. M. Carroll, V. Duvvuri, M. Trodden, M. S. Turner,
Phys. Rev. D {\bf 70}, 043528 (2004)
\bibitem{sm} S. M. Carroll,
A. De Felice, V. Duvvuri, D. A. Easson, M. Trodden
and M. S. Turner, Phys. Rev. D {\bf 71}, 063513 (2005)\\
G. Allemandi, A. Browiec and M. Francaviglia, Phys. Rev. D {\bf 70}, 103503 (2004)\\
X. Meng and P. Wang, Class. Quant. Grav. {\bf 21}, 951 (2004)\\
M. E. soussa and R. P. Woodard, Gen. Rel. Grav. {\bf 36}, 855
(2004)\\
S. Nojiri and S. D. Odintsov, Phys. Rev. D {\bf 68}, 123512 (2003)\\
P. F. Gonzalez-Diaz, Phys. Lett. B {\bf 481}, 353 (2000)\\
K. A. Milton, Grav. Cosmol. {\bf 9}, 66 (2003)
\bibitem{maeda} K. Maeda, Phys. Rev. D {\bf 39}, 3159 (1989)\\
 D. Wands, Class. Quant. Grav. {\bf 11}, 269 (1994)
\bibitem{soko} G. Magnano and L. M. Sokolowski, Phys. Rev. D {\bf 50}, 5039 (1994)
\bibitem{bis}Y. Bisabr, Phys. Lett. B {\bf 690}, 456 (2010)\\
Y. Bisabr, Phys. Rev. D {\bf 82}, 124041 (2010)
\bibitem{slow}R. J. Scherrer and A. A. Sen, Phys. Rev. D {\bf 77}, 083515
(2008)\\
R. J. Scherrer and A. A. Sen, Phys. Rev. D {\bf 78}, 067303
(2008)\\
S. Dutta, E. N. Saridakis and R. J. Scherrer, Phys. Rev. D {\bf
79}, 103504 (2009)\\
\bibitem{vik}A. Vikman, Phys. Rev. D {\bf 71}, 023515 (2005)
\bibitem{k}C. Armendariz-Picon, V. Mukhanov and
P. J. Steinhardt, Phys. Rev. D {\bf 63}, 103510 (2001)\\
A. Melchiorri, L. Mersini, C. J. Odman and M. Trodden, Phys. Rev.
D {\bf 68}, 043509 (2003)
\bibitem{multi}R. R. Caldwell and M. Doran, Phys.Rev. D {\bf 72}, 043527
( 2005)\\
W. Hu, Phys. Rev. D {\bf 71}, 047301 (2005)\\
Z. K. Guo, Y. S. Piao, X. M. Zhang and Y. Z. Zhang, Phys.Lett. B
{\bf 608}, 177 (2005)\\
B. Feng, X. L. Wang and X. M. Zhang, Phys. Lett. B {\bf 607}, 35
(2005)\\
B. Feng, M. Li, Y. S. Piao and X. Zhang, Phys. Lett. B {\bf 634},
101, (2006)
\bibitem{non}L. Perivolaropoulos, JCAP 0510, 001 (2005)
\bibitem{fa}V. Faraoni, E. Gunzig and P. Nardone, Fund. Cosmic Phys. {\bf 20}, 121 (1999)
\bibitem{dick}R. H. Dick, Phys. Rev. {\bf 125}, 2163 (1962)\\
J. D. Bekenstein, Phys. Rev. D {\bf 22}, 1313 (1980)\\
V. Faraoni, Cosmology in Scalar-Tensor Gravity, Kluwer Academic Publisher (2004)
\bibitem{20}V. Faraoni and E. Gunzig, Int. J. Theor. Phys. {\bf 38}, 217 (1999)\\
K. Nozari and S. D. Sadatian,  Mod. Phys. Lett. A {\bf 24}, 3143 (2009)\\
S. Capozziello, P. Martin-Moruno and C. Rubano,  Phys. Lett. B {\bf 689}, 117 (2010)
\bibitem{st}
S. Nojiri and S.D. Odintsov, Phys. Lett. B {\bf 652}, 343 (2007)\\
S. Nojiri and S.D. Odintsov, Phys. Lett. B {\bf 657}, 238 (2007)\\
L. Amendola, R. Gannouji, D. Polarski, S. Tsujikawa, Phys. Rev. D
{\bf 75}, 083504 (2007)\\
G. Cognola, E. Elizalde, S. Nojiri, S.D. Odintsov, L. Sebastiani,
S. Zerbini, Phys. Rev. D {\bf 77}, 046009 (2008)\\
A. Dev, D. Jain, S. Jhingan, S. Nojiri, M. Sami, I. Thongkool,
Phys. Rev. D {\bf 78}, 083515 (2008)\\
  T. P. Sotiriou and V.
Faraoni, Rev. Mod. Phys. {\bf 82}, 451 (2010)
\bibitem{dk}A. D. Dolgov, and M. Kawasaki, Phys. Lett. B {\bf
573},1 (2003)
\bibitem{wang}Y.G. Gong and A. Wang, Phys. Rev. D {\bf 73}, 083506
(2006)
\bibitem{sam}E. Komatsu et al., Astrophys. J. Suppl. {\bf 192}, 18 (2011)\\
M. Sullivan at al., Astrophys. J. {\bf 737}, 102 (2011)
\bibitem{q} Y. Gong and A. Wang, Phys. Rev. D {\bf 75}, 043520
(2007)
\bibitem{shap} C. A. Shapiro and M. S. Turner, astro-ph/0512586
\bibitem{cap}S. Capozziello, V. F. Cardone, S. Carloni and A. Troisi, Int. J. Mod. Phys. D {\bf 12}, 1969
(2003)
\bibitem{A}L. Amendola, R. Gannouji, D. Polarski and S. Tsujikawa, Phys.
Rev. D {\bf 75}, 083504 (2007)
\bibitem{star}A. A. Starobinsky, JETP. Lett. {\bf 86}, 157 (2007)
\bibitem{fla} E. E. Flanagan, Class. Quant. Grav. {\bf 21}, 3817 (2004)
\end{thebibliography}
\end{document}